\documentclass[aps,prd,showpacs,superscriptaddress,twocolumn,twoside,preprintnumbers]{revtex4} 
\usepackage{amssymb,amsmath}
\usepackage{graphicx,color} 
\usepackage{epsfig}
\begin{document}
\preprint{BI-TP 2015/14}

\title{Debye mass at the QCD transition in the PNJL model} 
\author{J.~Jankowski}
\email{jakubj@th.if.uj.edu.pl}
\affiliation{Institute of Physics, 
Jagiellonian University
ul. {\L}ojasiewicza 11 
30-348  Krak{\'o}w 
Poland}
\author{D.~Blaschke}
\email{blaschke@ift.uni.wroc.pl}
\affiliation{Institute for Theoretical Physics, University of Wroc{\l}aw, 
50-204 Wroc{\l}aw, Poland}
\affiliation{Bogoliubov Laboratory for Theoretical Physics, Joint Institute 
for Nuclear Research, RU - 141980 Dubna, Russia}
\author{O.~Kaczmarek}
\email{okacz@physik.uni-bielefeld.de}
\affiliation{Department of Physics, Bielefeld University,
D-33501 Bielefeld, Germany}

\date{\today} 

\begin{abstract}
We consider colour-electric screening as expressed by 
the quark contribution to the Debye mass
calculated in a PNJL model with emphasis on confining and
chiral symmetry breaking effects. 
We observe that the screening
mass is entirely determined by the nonperturbative quark
distribution function and temperature dependent QCD running coupling.
The role of the gluon background (Polyakov loop) is to provide strong
suppression of the number of charge carriers below the transition
temperature, as an effect of confinement, while the
temperature dependent dynamical quark mass contributes additional
suppression, as an effect of chiral symmetry breaking. 
An alternative derivation of this result from a modified kinetic theory is given,
which allows for a slight generalization and explicit contact with perturbative QCD. 
This gives the possibility to gain insights into the
colour screening mechanism in the region near the QCD pseudocritical temperature
and to provide a guideline for the interpretation of lattice QCD data.
\end{abstract}


\maketitle


\section{Introduction}
\label{Intro}

\par 
Theoretical and experimental investigations of quantum chromodynamics (QCD) 
at finite temperatures are performed with the aim to gain insights to the 
mechanisms of chiral symmetry restoration and deconfinement. 
From this perspective the heavy-quark (HQ) potential is one of the most important 
probes to be studied. 
In the vacuum the heavy quark-antiquark system is well described
in terms of effective field theories, owing to the 
energy scale separation \cite{Brambilla:1999xf,Brambilla:2004jw} 
and the fact that the HQ potential can be 
properly defined in terms of large Wilson
loops.
Then the spectrum of charmonium and bottomonium states as solutions of the 
Schr\"odinger equation for  heavy quarkonia can be extracted and faced experiment.
Finite temperature studies rely on the extraction of a static potential from 
{\it ab initio} simulations of lattice QCD (lQCD)
considering the singlet free energy due to a pair of static color charges as 
a function of their distance by means of Polyakov loop correlators 
\cite{Kaczmarek:2005ui}.
However, the systematic field theoretic description within the NRQCD framework 
\cite{Brambilla:2004jw} is much more subtle and delicate 
because new variable energy scales like 
the temperature $T$ or the screening mass $m_D$ enter the problem distorting
the energy hierarchy.
In addition, from the definition of a real-time potential \cite{Laine:2006ns} it
was realized that also an imaginary part in the potential appears at finite
temperature \cite{Laine:2007qy,Beraudo:2007ky,Brambilla:2008cx}. Using lattice
QCD studies of this complex values static potential it
was recently found that the real part is well described by the color singlet
free energy of a static quark-antiquark pair \cite{Burnier:2014ssa}.


While those results shed light on the important question 
about the identification of the HQ potential with
the colour singlet free energy \cite{Wong:2004zr}
another important question is that of microphysics insights into
the screening mechanism.
Based on some first principle motivations \cite{Megias:2007pq},   
Riek and Rapp have proposed an ansatz for the HQ potential \cite{Riek:2010fk}
in the form of a screened Cornell potential where the Coulomb and the linear parts
are subject to two different screening masses $m_D$ and $\tilde {m}_D$, respectively.
The fit of the temperature dependence of these parameters provided in Ref.~\cite{Riek:2010fk} 
using available lQCD data has revealed some unexpected aspects.
First, the coulombic Debye mass $m_D$ has a linear behaviour with
very small slope (smaller than expected from pQCD). 
Second, the screening mass of the confining part, $\tilde {m}_D$, shows a strong suppression for temperatures below $T_c$ and a linear rise for high temperatures (higher than expected from pQCD). 

This is somehow different from the standard approach,
where the lQCD data for the large distance part of this
HQ potential are fitted either to a Debye screened Coulomb ansatz
or to a form motivated by a Debye-H\"uckel theory
\cite{Kaczmarek:2007pb,Digal:2005ht,Burnier:2015nsa}.
An ansatz taking
into account real and imaginary parts of the HQ potential
has been recently considered in \cite{Burnier:2015nsa} in the context
of quenched lattice data.
Here we use $m_D(T)$ from lQCD data with
dynamical fermions obtained from fitting the heavy quark-antiquark free 
energy at large separations to
the standard Debye screened potential \cite{Kaczmarek:2007pb}
to compare with the model predictions.
While its behaviour well above the pseudocritical temperature $T_c$ of the QCD phase 
transition can qualitatively be understood in terms of perturbation theory, the 
interpretation of the lattice data in the vicinity of $T_c$ require 
essentially nonperturbative approaches addressing effects of confinement
and chiral symmetry breaking.  
The leading order perturbative result reads
$m_D=(1+\frac{N_f}{6})^\frac{1}{2}gT$, 
while the next-to-leading order can be obtained by 
resummation of the leading contribution of the high temperature expansion 
\cite{Rebhan:1993az,Rebhan:1994mx,Braaten:1994pk}.
A detailed understanding of the physics behind the Debye mass in the 
nonperturbative domain is subject to many current studies. 

Approaches based on the operator product expansion (OPE) 
\cite{Chakraborty:2011uw}, gauge/gravity duality \cite{Bak:2007fk,Finazzo:2014zga} 
or phenomenological models \cite{Megias:2007pq}
make an attempt to give a microscopic description of the screening phenomenon.
However, the very definition and numerical determination of a screening 
mass is obscured by the complications of the non-abelian
nature of QCD and the large value of the coupling constant near the QCD 
transition region \cite{Nadkarni:1986cz,Arnold:1995bh}. 

In this paper we investigate screening effects in a PNJL model which proved 
successful in reproducing various aspects of hadronic excitations in the medium
\cite{Hansen:2006ee} and of lQCD thermodynamics \cite{Ratti:2005jh}. 
We will evaluate the one-loop polarization function using PNJL propagators with QED 
like vertices thus extending a previous calculation made for massless quarks 
\cite{Jankowski:2009kr}.
In this rough way we implement confinement and chiral symmetry breaking 
effects which in turn allows a comparison to the lattice data for the Debye mass. 
We will show that one can reproduce the correct shape of its temperature dependence.
However, due to the absence of dynamical gluons in our PNJL model calculation,
we lack dynamical degrees of freedom and therefore stay below the lQCD result.

It is well known that the screening mass in QED or pQCD can also be derived from
kinetic theory \cite{Mrowczynski:1998nc,Blaizot:2001nr}. 
Interrelations between plasma physics and quark gluon plasma
are known to bring many relevant insights and physical motivation
behind the field theory calculations \cite{Mrowczynski:2007hb}.
We also explore this possibility and modify the standard kinetic theory 
approach by replacing usual Fermi-Dirac distribution functions for quarks with 
those modified by the coupling to the Polyakov loop. 
In this way we are able to reproduce the result for the Debye mass calculated within our model 
and furthermore to achieve contact with QCD by inclusion of effects of perturbative non-Abelian vertices.

The paper is organized as follows. 
In section \ref{model_mD} we outline the model calculation of the Debye mass within the PNJL model, whereby details are referred to the Appendix. 
Section \ref{KinTheo} presents the kinetic theory approach
to the problem where we give a simple and intuitive derivation of
the screening mass suggesting a straightforward modification of the standard approach.
Section \ref{LQCD} is devoted to a comparison with lQCD data
and their interpretation while section \ref{Concl} gives the conclusions.


\section{PNJL model calculation of the Debye mass}
\label{model_mD}



\begin{figure}[!htb]
\includegraphics[width=0.49\textwidth]{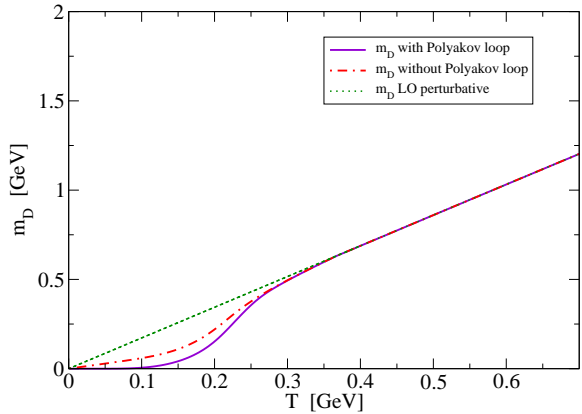}
\caption{Debye mass in leading order perturbation theory (dotted line),
including the effect of dynamical chiral symmetry breaking without coupling to the Polyakov loop
(dash-dotted line) and including the coupling to the Polyakov loop (solid line). 
Calculated with $\alpha_s=0.471$ fitted to charmonium spectrum at $T=0$. }
\label{Fig.1}
\end{figure}

In reference \cite{Jankowski:2009kr} a model was considered 
where the vacuum HQ potential was screened by a quark loop with
internal lines coupled to a temporal background gluon field.
For the static interaction potential $V(q)$, $ q^{2} = |{\bf{q}}|^{2} $, 
the statically screened potential is given by a resummation of 
one-particle irreducible diagrams (RPA "bubble" resummation)
\begin{equation}
V_{\rm sc}(q) = {V(q)}/[{1 + F(0; {\bf q})/q^{2}}]~,
\label{Vsc}
\end{equation}
where the longitudinal polarization function  
in the finite temperature case is defined via the 
projector decomposition of the self energy
(in Euclidean space)
\begin{equation}
\Pi_{\mu\nu}(i\omega,q) = F(i\omega,q) P^L_{\mu\nu} + G(i\omega,q)P^T_{\mu\nu}~,
\label{eq:Decomposition}
\end{equation}
where the projectors satisfy
\begin{equation}
(P^L)^2=(P^T)^2=1 \hspace{15pt} P^TP^L=P^LP^T=0~,
\label{eq:}
\end{equation}
\begin{equation}
P^L_{\mu\nu} + P^T_{\mu\nu} = \delta_{\mu\nu} - \frac{q_\mu q_\nu}{Q^2}~,
\label{eq:}
\end{equation}
\begin{equation}
P^T_{ij} = \delta_{ij} - \frac{Q_iQ_j}{Q^2} \hspace{15pt} P^T_{44}=P^T_{j4}=P^T_{4j}=0~.
\label{eq:}
\end{equation}
Here, $Q=(\omega,{\bf q})$ is the Euclidean four momentum. 
From this we get the gauge invariant longitudinal component
\begin{equation}
F(i\omega,q) = \frac{Q^2}{q^2}\Pi_{44}(i\omega,q)~,
\label{eq:}
\end{equation}
and it can be calculated within thermal field theory as
\begin{widetext}
\begin{equation}
\Pi_{44}(i\omega_{l};{\bf q} ) 
= g^{2} T\sum_{n=-\infty}^{\infty} \int\frac{d^{3}p}{(2\pi)^{3}} 
{\textrm{ Tr}} [\gamma_{4}S_{\Phi}(i\omega_{n}; {\bf p})
\gamma_{4}S_{\Phi}(i\omega_{n}-i\omega_{l}; {\bf p} - {\bf q})]~.
\end{equation}
\end{widetext}
Here $\omega_{l}=2\pi lT$ are the bosonic and $\omega_{n}=(2n+1)\pi T$
the fermionic Matsubara frequencies of the imaginary-time formalism.
The symbol ${\textrm{Tr}}$ stands for traces in color, flavor and Dirac 
spaces.
$S_{\Phi}$ is the fermionic quark propagator coupled to the homogeneous 
static gluon background field $\varphi_3$. Its inverse is given by 
\cite{Ratti:2005jh,Hansen:2006ee}
\begin{equation}
S^{-1}_{\Phi}( {\bf p}; i\omega_{n} ) = 
{\bf \gamma\cdot p} - \gamma_{4}\omega_{n} + \gamma_4\lambda_{3}\varphi_3+m~,
\label{coupling}
\end{equation}
where $\{\gamma_\mu,\gamma_\nu\}=-2\delta_{\mu\nu}$ and $m=m(T)$ is the 
dynamically generated temperature dependent mass for light quarks as described,
e.g., within the NJL model \cite{Klevansky:1992qe}.
The variable $\varphi_3$ is related to the Polyakov loop variable defined by
\cite{Ratti:2005jh}
$$ \Phi(T) = \frac{1}{3}\rm Tr_c (e^{i\beta\lambda_{3}\varphi_{3}}) 
= \frac{1}{3}(1 + 2\cos(\beta\varphi_3) )~, $$ 
The physics of $\Phi(T)$ is governed by the temperature-dependent Polyakov 
loop potential ${\cal{U}}(\Phi)$, which is fitted to describe the lattice data 
for the pressure of the pure glue system  \cite{Ratti:2005jh}. 
After performing the color-, flavor- and Dirac traces and making the fermionic 
Matsubara summation 
we obtain \cite{Kapusta:2006pm} 
(see Appendix \ref{App} for the details)
\begin{eqnarray}
\Pi_{44}(i\omega,q) &=& g^2\textrm{Re}\int_0^\infty \frac{p^2dp}{\pi^2} \frac{2f_\Phi(E_p)}{E_p}
\nonumber\\
&&\Bigg\{1 + \frac{4E_pi\omega + q^2+\omega^2-4E_p^2}{4pq}\ln\frac{R_+(\omega)}{R_-(\omega)}\Bigg\}~,
\nonumber\\
\label{eq:}
\end{eqnarray}
where
\begin{equation}
R_\pm(\omega) = -\omega^2-q^2-2i\omega E_p\pm2pq~,
\label{eq:}
\end{equation}
and $\textrm{Re}f(\omega):=\frac{1}{2}(f(\omega)+f(-\omega))$. 
Taking the static, long wavelength limit 
\cite{Kapusta:2006pm,LeBellac,Hansen:2006ee}
we identify, after continuation $i\omega\rightarrow q_0+i\epsilon$
to the Minkowski space,
$F(q_0=0,q\rightarrow0)=m_D^2(T)$
and whence
\begin{eqnarray}
m_{D}^2(T) = \frac{16\alpha_s}{\pi} \int_{0}^{\infty}dp\, 
\left[p^{2} + E_p^2\right] \frac{f_\Phi(E_p)}{E_p} ~.
\label{debyemass}
\end{eqnarray}
Here $m_{D}(T)$ is the Debye mass, $E_p=\sqrt{p^2+m^2}$ is the quasiparticle
dispersion relation for light quarks with $N_c=3$ colour and $N_f=2$ flavour 
degrees of freedom;
$f_\Phi(E)$ is the quark distribution function \cite{Hansen:2006ee}
\begin{equation}
f_\Phi(E) = 3\frac{\Phi(1+2e^{- \beta E})e^{- \beta E}+e^{-3 \beta E}}
{1 + 3\Phi(1 + e^{- \beta E})e^{- \beta E}+e^{-3 \beta E}}~. 
\label{eq:Dist}
\end{equation} 
This result is different from the standard QED case only in that the 
Fermi-Dirac distribution has been replaced by the function (\ref{eq:Dist}).
In comparison to the free fermion case \cite{LeBellac,Beraudo:2007ky} the 
coupling to the Polyakov loop variable $\Phi(T)$ gives rise to a modification 
of the Debye mass, encoded in the modification of usual Fermi-Dirac
distribution function. 
In the limit of deconfinement ($\Phi = 1$), the case of a massive
quark gas is obtained, while for confinement ($\Phi = 0$) one finds
a considerable suppression. 
The temperature dependence of the Debye mass is shown in figure~\ref{Fig.1} 
and as expected from the very beginning it turns out to be below the free massless case, 
in the confined and in the transition region (with a pseudocritical temperature 
$T_c\approx200~\textrm{MeV}$).
For temperatures $T>>T_c$ the free gas behaviour $m_D^2=2/3g^2T^2$ is 
reproduced.

The above result can be also obtained in a different way.
It is well known that in QED the Debye mass is related to the 
pressure via the second derivative \cite{Kapusta:2006pm,LeBellac}  
\begin{equation}
m_D^2(T,\mu_e) = e^2 \frac{\partial^2 P(T,\mu_e)}{\partial\mu_e^2}~,
\label{eq:mDQED}
\end{equation}
where here $\mu_e$ is related to the electric charge of the system. 
This relation is a consequence of the Dyson-Schwinger 
equation for the photon self energy and the Ward identity relating the
electron-photon vertex to the quark propagator. 
This is only true in abelian gauge theory and breaks down for non-abelian theories
like QCD. 
On the technical level the proper Ward identity (called then Slavnov-Taylor identity) 
becomes much more complex and does not allow a simple derivation.
Because our calculation in this section is similar to
a one-loop QED calculation it is interesting to see if a similar
relation holds also in the PNJL model. We check this in the 
finite temperature and zero chemical potential case by 
directly evaluating right hand side of eq. (\ref{eq:mDQED}). 
Let us recall the quark contribution to the meanfield pressure 
reads \cite{{Hansen:2006ee}}
\begin{eqnarray}
P_q(T,\mu) &=& -2N_f T \int \frac{d^3p}{(2\pi)^3}
\nonumber\\
&&\left\{\ln\left[1+3(\bar \Phi +\Phi X_-)X_- +X_-^3\right]
\right.
\nonumber\\
&&
+\left. \ln\left[1+3(\Phi +\bar \Phi X_+)X_+ + X_+^3\right]\right\},
\label{eq:Pq}
\end{eqnarray}
where $X_\mp=e^{-\beta (E_p\mp \mu)}$.
The vacuum pressure has been subtracted and at
finite $\mu$ the function $\Phi$ is generally different from its
complex conjugate $\bar \Phi$.
In the small density limit constituent quark mass 
and expectation value of traced Polyakov loop are 
$\mu$ independent so the second derivative of the 
quark  pressure simplifies giving after noticing
that it can be written as derivative with respect
to the quasiparticle energy $E_p$
\begin{eqnarray}
m_D^2(T) &=& -g^2 \frac{\partial^2 P}{\partial\mu^2}(T,\mu=0) 
\nonumber\\
&=&
12 g^2 N_f \int \frac{d^3p}{(2\pi)^3} \frac{d}{dE_p}\left[ \frac{X^3+\Phi(X+2X^2)}{1+3\Phi(X+X^2)+X^3}\right],
\nonumber\\
\label{eq:}
\end{eqnarray}
where we have used $X=e^{-\beta E_p}$ and the fact that $\Phi=\bar \Phi$
for $\mu=0$. 
The quantity under the integral can be 
identified with the energy derivative of our modified
distribution function bringing us to the following 
formula
\begin{equation}
m_D^2(T) = -\alpha_s \frac{8N_f}{\pi} \int_0^\infty dp p^2 \frac{df_\Phi}{dE_p}(E_p)~,
\label{eq:}
\end{equation}
which after integration by parts (see section \ref{KinTheo})
gives for $N_f=2$ the result (\ref{debyemass}).


\section{Kinetic theory approach}
\label{KinTheo}


The usual Debye screening mass in QED
or perturbative QCD can be derived within a kinetic theory approach 
\cite{Yagi:2005yb,Mrowczynski:1998nc}.
In this section we will modify the standard kinetic theory 
so that we consistently reproduce our previous Debye mass derivation 
and generalize it in order to make contact with perturbative 
calculations for high temperatures. 
The kinetic theory approach has been widely used in the context of perturbative QCD
\cite{Blaizot:2001nr,Litim:1999id}
providing a physical picture behind the hard thermal loop
approximation and some insights into transport properties and
collective modes of the quark gluon plasma.
The appropriate change with respect to the textbook result is that 
the usual Fermi Dirac distribution function is replaced by the 
Polyakov loop modified distribution function (\ref{eq:Dist}).
Then the charge density induced by the electrostatic potential
$A_0(x)=V(x)$ can be written
\begin{eqnarray}
\rho_{\textrm{ind}}(x) &=& 2g\int \frac{d^3 p}{(2\pi)^3} \left[f_\Phi(E_p+gV(x))-f_\Phi(E_p-gV(x))\right]
\nonumber\\
&&\approx 4g^2\int\frac{d^3 p}{(2\pi)^3} \frac{df_\Phi}{dE_p}(E_p)V(x)~,
\label{eq:}
\end{eqnarray} 
where the factor $2$ is due to the fermion spin. 
The Maxwell equations give
\begin{eqnarray}
-\nabla^2 V(x) &=& \rho_{\textrm{ind}}(x) 
= \frac{2g^2}{\pi^2}\int_0^\infty dp p^2 \frac{df_\Phi}{dE_p}(E_p)V(x) \nonumber\\
&=& - m_D^2(T)V(x)~, 
\label{eq:Maxwell}
\end{eqnarray}
where 
\begin{eqnarray}
m_D^2(T) &=& -\frac{2N_fg^2}{\pi^2}\int_0^\infty dp p^2 \frac{df_\Phi}{dE_p}(E_p)
\nonumber\\
&=& -\frac{8N_f\alpha_s}{\pi} \int_{0}^{\infty}dp\, 
p\sqrt{p^{2} + m^2} \frac{df_\Phi(E_p)}{dp} ~,
\label{eq:}
\end{eqnarray}
and the relation $E_pdE_p=pdp$ has been used. 
Integrating by parts,
\begin{eqnarray}
m_D^2(T) &=& -\frac{8N_f\alpha_s}{\pi} \int_{0}^{\infty}dp\, 
p\sqrt{p^{2} + m^2} \frac{df_\Phi(E_p)}{dp} 
\nonumber\\
&=&-\frac{8N_f\alpha_s}{\pi}
\left[p\sqrt{p^2+m^2}f_\Phi(E_p)|_{p=0}^{p=\infty} \right.
\nonumber\\ 
&& \left.- \int_{0}^{\infty}dp f_\Phi(E_p) \frac{d}{dp}
\left(p\sqrt{p^2+m^2}\right)\right]~,
\label{eq:}
\end{eqnarray}
one arrives at the result for the Debye screening mass for $N_f$ flavours of fermions
\begin{equation}
m_D^2(T) 
= \frac{8N_f\alpha_s}{\pi} \int_{0}^{\infty}dp\, 
\left[p^{2} + E_p^2\right] \frac{f_\Phi(E_p)}{E_p} ~.
\label{eq:}
\end{equation}
For the non-abelian case the induced density reads
\begin{equation}
\rho_{\textrm{ind}}(x) = 2g\int \frac{d^3 p}{(2\pi)^3} 
\left[f_+^b(p,x)-f_-^b(p,x)\right]\textrm{Tr}[t^bt^a]~, 
\label{eq:}
\end{equation}
where $\textrm{Tr}[t^bt^a]=\frac{1}{2}\delta^{ab}$ and we assume
\begin{equation}
f_\pm^a(x,p)=\pm g \frac{df_\Phi}{dE_p}(E_p)V^a(x)~. 
\label{eq:}
\end{equation}
Doing the same steps as before will give the Debye mass ($N_f=2$)
\begin{equation}
m_D^{*2}(T) = \frac{8\alpha_s}{\pi} \int_{0}^{\infty}dp\, 
\left[p^{2} + E_p^2\right] \frac{f_\Phi(E_p)}{E_p} ~,
\label{eq:mD_QCD}
\end{equation} 
which now reproduces the Debye mass of perturbative QCD  for high temperatures.


\section{Comparison with Lattice QCD}
\label{LQCD}


Within lQCD, the temperature dependent
screening masses have been defined from the
exponential fall-off of the colour singlet
free energies \cite{Kaczmarek:2005ui} and
the results could be represented in the rescaled
leading order perturbative result
\begin{equation}
\frac{m_D(T)}{T} = A(T)\left(1+\frac{N_f}{6}\right)^{1/2} 
g_{2-\textrm{loop}}(T)~,
\label{eq:}
\end{equation}
where the factor $A(T)$ was introduced to account for nonperturbative effects.
The analysis performed in \cite{Kaczmarek:2007pb} has shown that 
$A(T) \approx 1.66$ for $N_f=2+1$ and $T \geq 1.2~T_c$.
Obviously, $A(T) \rightarrow 1$ for high temperatures in agreement with 
perturbation theory.
In order to compare our model with lQCD data we have to adopt  equation 
(\ref{eq:mD_QCD}),
because it makes contact with perturbation theory
for high temperatures. 
The running coupling constant is modelled in two forms.
The first one being the two-loop result
\cite{Caswell:1974gg}
\begin{equation}
\alpha(T) = \frac{1}{4\pi\left[2\beta_0\ln(\frac{\mu T}{\Lambda})+(\frac{\beta_1}{\beta_0})\ln(2\ln(\frac{\mu T}{\Lambda}))\right]}~,
\label{eq:}
\end{equation}
where $\mu=\pi$ is the upper
bound for the perturbative coupling.
For $\Lambda$ we use a value in the ${\overline{\rm MS}}$ scheme of 
$\Lambda_{\overline{MS}}=260$~MeV   
and
\begin{equation}
\beta_0 = \frac{1}{16\pi^2}\left(11-\frac{2N_f}{3}\right)~,
\label{eq:}
\end{equation}
\begin{equation}
\beta_1 = \frac{1}{(16\pi^2)^2}\left(102 - \frac{38N_f}{3}\right)~.
\label{eq:}
\end{equation}
The other possibility is to use a running coupling constant obtained
by solving the one-loop renormalization group equation with pole subtraction 
\cite{Shirkov:1997wi}
\begin{equation}
\alpha_s(T) = 
\frac{4\pi}{\widetilde{\beta}_0}\left[\frac{1}{2\ln\frac{\pi T}{\Lambda}} 
+ \frac{\Lambda^2}{\Lambda^2-(\pi T)^2}\right]~,
\label{eq:}
\end{equation}
where $\widetilde{\beta}_0=11-2/3N_f$.

In this way we can identify our model predictions
for the nonperturbative effects of confinement
and chiral symmetry restoration which are 
expressed as a temperature dependent
factor $A(T)$ which reads (for $N_f=2$)
\begin{equation}
A^2(T) = \frac{6}{\pi T^2}\int_0^\infty~dp\frac{f_\Phi(E_p)}{E_p}
\left\{p^2 + E_p^2\right\}~.
\label{eq:}
\end{equation} 
We see that it is entirely controlled by the quark distribution function 
and by the temperature dependent quark mass which mimic confinement as well
as chiral symmetry breaking aspects of strong interaction dynamics.

For a comparison to lattice QCD data we have chosen the $N_t=6$ data for
$2+1$-flavors from \cite{Kaczmarek:2007pb}.
There two definitions of the running coupling were used, $\alpha_s$ from a fit using a Debye screened Coloumb Ansatz at large separations and $\alpha_{\rm max}$ obtained at the maximum of the effective $r$ 
and $T$-dependent running coupling.
Those results are based on an analysis of
gauge field configurations generated by the RBC-Bielefeld collaboration in (2+1)-flavor QCD for
the calculation of the QCD equation of state \cite{Cheng:2007jq} where the pion
mass is about 220 MeV and the strange
quark mass is adjusted to its physical value.

\begin{figure}[!h]
\includegraphics[width=0.49\textwidth]{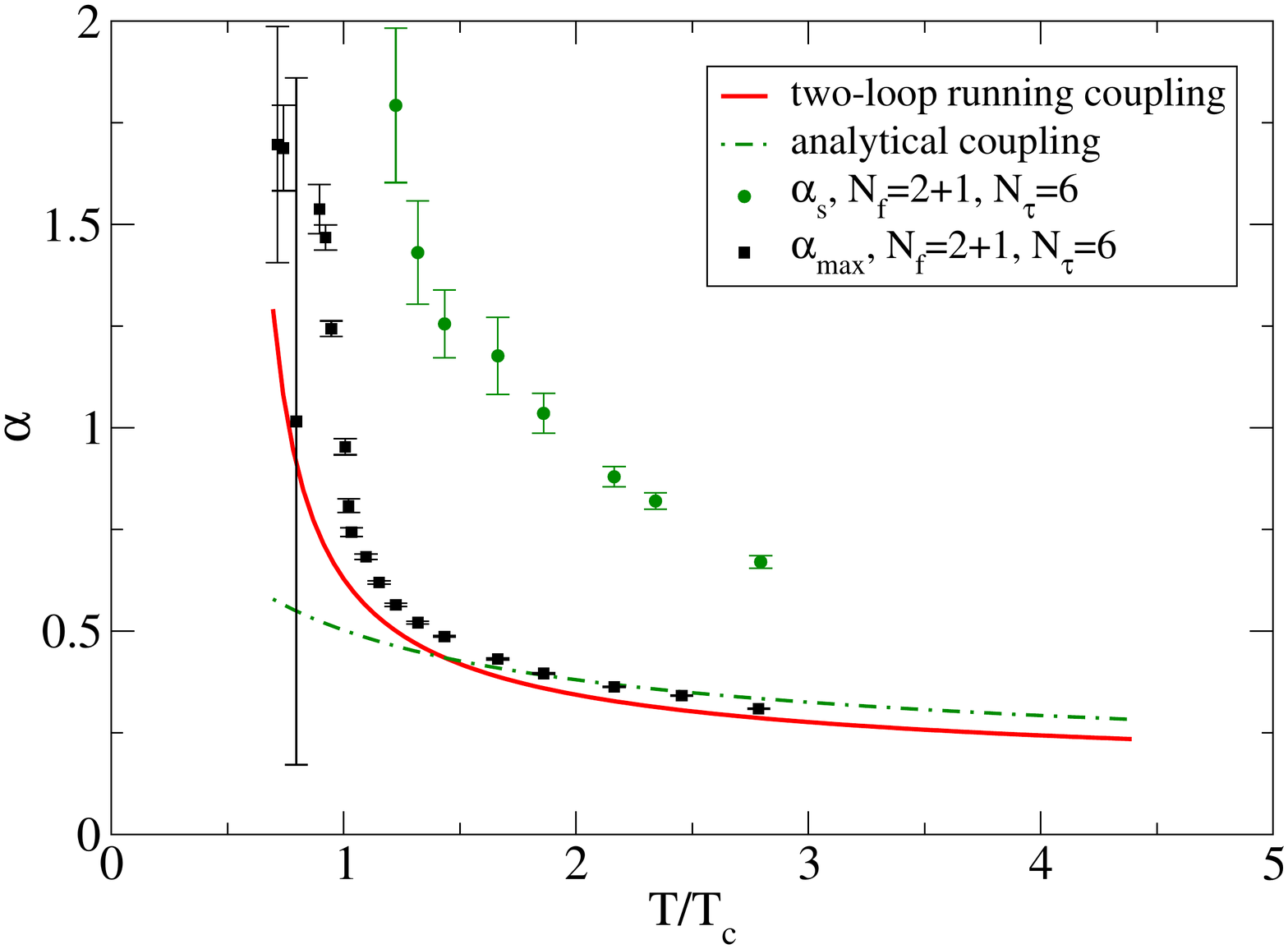}
\includegraphics[width=0.49\textwidth]{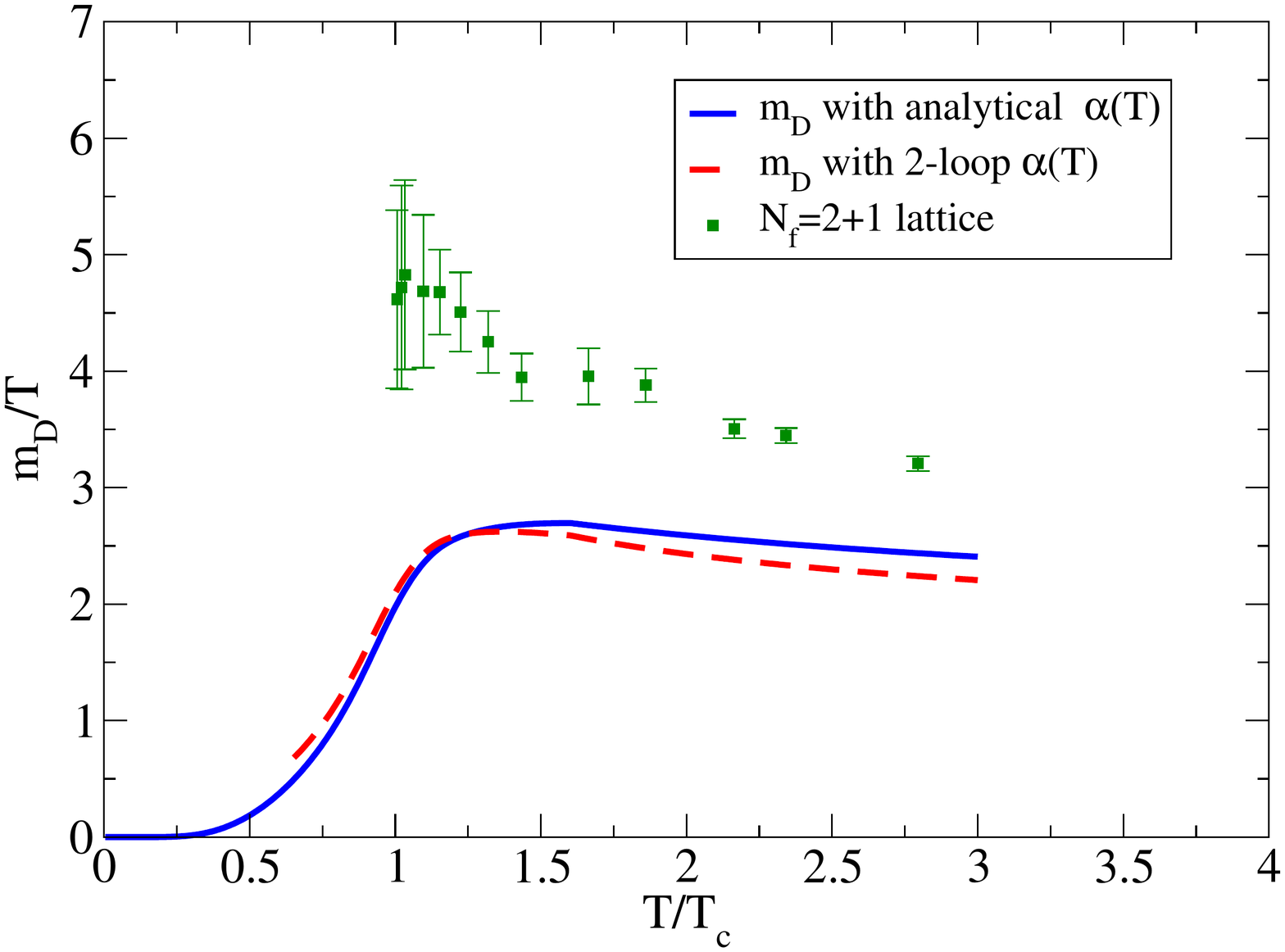}
\caption{
Upper panel: Two loop running coupling an the analytical 
running coupling \cite{Caswell:1974gg,Shirkov:1997wi}. 
Lower panel: Debye mass with coupling to the Polyakov loop and
running coupling constant $\alpha(T)$ compared
with the lattice data \cite{Kaczmarek:2005ui}. }
\label{Fig.2}
\end{figure}

From the lower panel of Fig.~\ref{Fig.2} we see that the general trend of 
the lattice data is reproduced. 
The fact that our result for the Debye mass stays below the lattice data
is due to the lack of dynamical gluons which in full QCD also bring a 
contribution to the screening effects. 
Also the smoother behaviour around $T_c$ is due to the used form of the 
running coupling which in principle is not applicable in the transition 
region. 
Below $T_c$ the suppression of  colour charges (which is meant as a rough form 
of confinement) drives $m_D(T)$ fastly to zero and overcompensates the increasing 
interaction strength which taken alone would tend to increase the screening mass.
Note that the quenched results for $m_D$ in \cite{Kaczmarek:2007pb} and \cite{Burnier:2015nsa} 
lie below the lattice data shown in Fig.~\ref{Fig.2} due to the missing degrees of freedom dynamical quarks.

\section{Conclusions}
\label{Concl}

In this short note we have presented a calculation of the quark contribution 
to the Debye screening mass using the PNJL model which 
captures chiral dynamics and in a very rough way some aspects of confinement. 
We have compared our results to the Debye masses extracted from 
lattice QCD simulations of the static heavy-quark potential and obtain overall 
agreement for the shape of the temperature dependence. 
Naturally, our results for the Debye mass stay below the lattice results since 
in our model the gluon contribution is neglected. 
However, as the observed gluon contribution on the lattice is of the same shape 
\cite{Kaczmarek:2007pb} we are convinced that our model captures the essentials 
of the influence of chiral dynamics and confinement on the screening of the 
heavy-quark potential. 
A further improvement of the calculation would be to include dynamical gluons into the system,
to improve the modelling of quark (and gluon) confinement
and to elaborate on the behaviour of the running coupling constant. 
The latter should also be compared with the lattice QCD result for this quantity as it is obtained simultaneously with that for the Debye mass.
At this point it is interesting to go back to the  different interpretations of the same lattice data. 
Here we would like to mention the one by Ref.~\cite{Riek:2010fk} 
where the ansatz of a screened Corrnell type potential was adopted with two 
different Debye masses, one for the linear confining part ($\widetilde{m}_D$)
and one for the Coulombic part ($m_D$).
The performed fit gave a drastically different behaviour of the two screening 
masses.
The temperature dependence obtained for $\widetilde{m}_D$ appears similar to
that of the Debye mass in the present approach, calculated for a $T$- 
independent coupling constant (see Fig.~\ref{Fig.1}).
The physical reason for such a distinction could be that the stringy and the 
Coulombic parts of the potential act on different length scales so that the
screening of them involves different dynamics. 
The linear part should be dominant for larger distances thus 
involving stronger interactions and more correlations in the screening 
mechanism. 
Thus one could expect that $\widetilde{m}_D>m_D$ for all temperatures which is 
the finding of  \cite{Riek:2010fk}.
This is, however, only a qualitative argument. 
Our calculation, since it is at one-loop order with a QED like
interaction should apply to the screened Coulomb potential part from 
which the lattice QCD result has been extracted.
As has been demonstrated here, this comparison provides a reasonable 
interpretation of the temperature dependence of the Debye mass.

\begin{acknowledgements}
We thank Alexander Rothkopf and Yannis Burnier for their detailed comments on
the first version of the manuscript.
We thank Rafa{\l} {\L}astowiecki for providing data for the PNJL model. 
Edwin Laermann and Krzysztof Redlich contributed with their discussions.
This work has been supported in part by NCN under grant number UMO-2011/02/A/ST2/00306.
The work of J.J. was partially supported by the NCN post-doctoral
internship grant DEC-2013/08/S/ST2/00547.
D.B. and J.J. gratefully acknowledge the hospitality of the University of Bielefeld extended to them 
during their visits, while O.K. is grateful for the hospitality and productive atmosphere in the Particle 
Physics group at the Institute for Theoretical Physics of the University of Wroclaw. 
\end{acknowledgements}


\onecolumngrid

\begin{appendix}
\section{Polarization loop, temporal component}
\label{App}

The calculation we have performed follows
in all steps the standard QED evaluation
of the polarization loop \cite{LeBellac,Kapusta:2006pm}
with the only difference that the usual Matsubara
summation is now equipped with a trace over the 
colour indices \cite{Hansen:2006ee}.
To start with let us define the  $44$ component
of the polarization tensor
\begin{equation}
\Pi_{44}=g^2T\sum_n\int \frac{d^3p}{(2\pi)^3}\textrm{Tr}\Bigg\{\left[\gamma_4(m+\gamma_4\omega_n-\vec{\gamma}\vec{p})\gamma_4(m+\gamma_4\omega_n-\gamma_4\omega_l-\vec{\gamma}(\vec{p}-\vec{q}))\right]
\Delta(i\omega_n,\vec{p})\Delta(i\omega_n-i\omega_l,\vec{p}-\vec{q})\Bigg\}~,
\end{equation}
where Tr stands for the trace in Dirac and color spaces, 
$\omega_n=(2n+1)\pi T - A_4$, 
with the temporal gluon field $A_4$, $\omega_l=2\pi lT$.
Let us define
\begin{equation}
\Delta(i\omega_n,\vec{p}) = \frac{1}{\omega_n^2+p^2+m^2} = \sum_{s=\pm} \frac{s}{2E_p} \frac{1}{i\omega_n+sE_p}~,
\label{eq:}
\end{equation} 
where $E_p=\sqrt{p^2+m^2}$. 
The first step is to calculate the Dirac trace using $\{\gamma_\mu,\gamma_\nu\}=-2\delta_{\mu\nu}$
\begin{equation}
\textrm{Tr}\left[(m+\gamma_4\omega_n+\vec{\gamma}\vec{p})\gamma_4^2(m+\gamma_4\omega_n-\gamma_4\omega_l-\vec{\gamma}(\vec{p}-\vec{q}))\right] =
-4\left[m^2-\omega_n(\omega_n-\omega_l)+\vec{p}(\vec{p}-\vec{q})\right]~,
\label{eq:}
\end{equation}
\begin{equation}
\mathcal{N}_V = \omega_n(\omega_n-\omega_l)-\vec{p}(\vec{p}-\vec{q})-m^2 = \omega_n^2-\omega_n\omega_l-p^2+pq-m^2~, 
\label{eq:}
\end{equation}
\begin{equation}
\textrm{Tr}\left[(m+\gamma_4\omega_n+\vec{\gamma}\vec{p})\gamma_4^2(m+\gamma_4\omega_n-\gamma_4\omega_l-\vec{\gamma}(\vec{p}-\vec{q}))\right] =
4\mathcal{N}_V~.
\label{eq:}
\end{equation}
\begin{eqnarray}
\Pi_{44}(i\omega_{l};{\bf q} ) 
= \frac{g^2T}{2}
T\sum_{n=-\infty}^{\infty} \int\frac{d^{3}p}{(2\pi)^{3}} 
\textrm{Tr}\Bigg\{2\Delta(i\omega_n,\vec{p})
+(4pq-4E_p^2-q^2-\omega_l^2)\Delta(i\omega_n,\vec{p})
\Delta(i\omega_n-\omega_l,\vec{p}-\vec{q})\Bigg\}~.
\label{eq:}
\end{eqnarray} 
Now we want to decompose (\ref{eq:}) so that we have 
$\Delta(i\omega_n,\vec{p})+\Delta(i\omega_n-i\omega_l,\vec{p}-\vec{q})$
and for that we introduce
\begin{equation}
\mathcal{M} = p^2+2m^2+\omega_n^2+(\omega_n-\omega_l)^2+(p-q)^2 = 2(p^2+m^2+\omega_n^2-\omega_n\omega_l-pq)+\omega_l^2+q^2~,
\label{eq:}
\end{equation} 
\begin{equation}
2\mathcal{N}_V = \mathcal{M} + 4pq -4p^2 - 4m^2 - q^2 -\omega_l^2~.
\label{eq:}
\end{equation}
Further, we evaluate three coloured Matsubara sums
\begin{equation}
J_3 = T\sum_n\textrm{Tr}_c\left[\Delta(i\omega_n,\vec{p})\Delta(i\omega_n-i\omega_l,\vec{p}-\vec{q})\right]~,
\label{eq:}
\end{equation}
\begin{equation}
J_1 = T\sum_n\textrm{Tr}_c\Delta(i\omega_n,\vec{p})~.
\label{eq:}
\end{equation}
\begin{equation}
J_2 = T\sum_n\textrm{Tr}_c\left\{[\vec{p}(\vec{p}-\vec{q})]\Delta(i\omega_n,\vec{p})\Delta(i\omega_n-i\omega_l,\vec{p}-\vec{q})\right\}~,
\label{eq:}
\end{equation}
following \cite{Hansen:2006ee} by going to the Polyakov gauge (where Polyakov
loop variable is diagonal) and explicitly calculating the trace. 
The outcome reads 
\begin{equation}
J_3 = \sum_{s,s'=\pm} \frac{ss'}{4E_pE_{p-q}}\frac{1}{i\omega+sE_p-s'E_{p-q}}\left[f_\Phi(sE_p)- f_\Phi(s'E_{p-q})\right]~,
\label{eq:}
\end{equation}
\begin{eqnarray}
J_3 = \frac{1}{4E_pE_{p-q}}\Bigg\{[f_\Phi(E_p)-f_\Phi(E_{p-q})]\left(\frac{1}{i\omega+E_p-E_{p-q}}-\frac{1}{i\omega-E_p+E_{p-q}}\right)+\\
\nonumber
[1-f_\Phi(E_p)-f_\Phi(E_{p-q})]\left(\frac{1}{i\omega+E_p+E_{p-q}}-\frac{1}{i\omega-E_p-E_{p-q}}\right)\Bigg\}~,
\label{eq:}
\end{eqnarray}
\begin{equation}
J_2 = \sum_{s,s'=\pm} \frac{ss'}{4E_pE_{p-q}}\frac{\vec{p}(\vec{p}-\vec{q})}{i\omega+sE_p-s'E_{p-q}}\left[f_\Phi(sE_p)-f_\Phi(s'E_{p-q})\right]~,
\label{eq:}
\end{equation}
\begin{eqnarray}
J_2 = \frac{\vec{p}(\vec{p}-\vec{q})}{4E_pE_{p-q}}\Bigg\{[f_\Phi(E_p)-f_\Phi(E_{p-q})]\left(\frac{1}{i\omega+E_p-E_{p-q}}-\frac{1}{i\omega-E_p+E_{p-q}}\right)+\\
\nonumber
[1-f_\Phi(E_p)-f_\Phi(E_{p-q})]\left(\frac{1}{i\omega+E_p+E_{p-q}}-\frac{1}{i\omega-E_p-E_{p-q}}\right)\Bigg\}~,
\label{eq:}
\end{eqnarray}
\begin{equation}
J_4 = \sum_{s=\pm}\frac{s}{2E_p}f_\Phi(-sE_p) = \frac{1}{2E_p}[1-2f_\Phi(E_p)]~,
\end{equation}
which is in agreement with (6.27) and (6.37) from LeBellac, 
with the only difference that Fermi-Dirac distribution functions
have been replaced with the modified ones  \cite{Hansen:2006ee}
\begin{equation}
f_\Phi(E) = T\sum_n\textrm{Tr}_c\left[\frac{1}{i\omega_n-E}\right] = 3\frac{\Phi(1+2e^{-\beta E})e^{-\beta E}+e^{-3\beta E}}{1+3\Phi(1+e^{-\beta E})e^{-\beta E}+e^{-3\beta E}}~.
\end{equation}
To obtain the last equation we use the fact that in this specific gauge the Polyakov loop variable is diagonal
and that after a Matsubara summation we get
\begin{equation}
f_\Phi(E) = \sum_{j=1}^3 \frac{1}{1+e^{\beta A_{jj}}e^{\beta E}}=-\frac{1}{\beta}\frac{\partial}{\partial E}\sum_{j=1}^3\ln(1+L_{jj}e^{-\beta E})~,
\label{eq:}
\end{equation}
where $L_{jj}=e^{-\beta A_{jj}}$ and $A$ 
is to be understood as a temporal component of the gauge field.
The evaluation of the colour trace is now trivial and results in
\begin{equation}
\sum_{j=1}^3\ln(1+L_{jj}e^{-\beta E})=\ln\left[(1+L_{11}e^{-\beta E})(1+L_{22}e^{-\beta E})(1+L_{33}e^{-\beta E})\right]~.
\label{eq:}
\end{equation}
Using $L_{11}+L_{22}+L_{33}=\textrm{Tr}L=\textrm{Tr}L^\dagger = 3\Phi$,
$L_{11}L_{22}L_{33}=\det L=\det L^\dagger = 1$ we get
\begin{equation}
f_\Phi(E) = -\frac{1}{\beta}\frac{\partial}{\partial E}\ln[1+3\Phi(1+e^{-\beta E})e^{-\beta E}+e^{-3\beta E}]~,
\label{eq:}
\end{equation}
which is the modified distribution function (\ref{eq:Dist}).
We also show that
\begin{equation}
f_\Phi(x+i\omega_l)=f_\Phi(x)~, \hspace{75pt} f_\Phi(-E)=1-f_\Phi(E)~.
\label{eq:}
\end{equation}
Now to get rid of the $f(E_{p-q})$ terms whenever we meet them we 
change the variables according to \cite{Klevansky:1992qe} $p\rightarrow -p+q$ 
so that $E_{p-q}\rightarrow E_p$, $E_p\rightarrow E_{p-q}$ and 
$\vec{p}(\vec{p}-\vec{q})$ does not change.
We also make the Wick rotation $i\omega\rightarrow\omega$
%
\begin{equation}
J_3 = \frac{f_\Phi(E_p)}{E_p}\left[\frac{1}{(\omega+E_p)^2-E_{p-q}^2} + \frac{1}{(\omega-E_p)^2-E_{p-q}^2}\right]~,
\label{eq:}
\end{equation}
\begin{equation}
J_2 = \vec{p}(\vec{p}-\vec{q})\frac{f_\Phi(E_p)}{E_p}\left[\frac{1}{(\omega+E_p)^2-E_{p-q}^2} + \frac{1}{(\omega-E_p)^2-E_{p-q}^2}\right]
=
\vec{p}(\vec{p}-\vec{q})J_3~.
\label{eq:}
\end{equation}
%
%
%
\begin{equation}
E_p^2 - E_{p-q}^2 = 2pq - q^2 \hspace{20pt} \vec{p}\vec{q}=pq\lambda \hspace{20pt} 
\int \frac{d^3p}{(2\pi)^3} = \int \frac{dpp^2 d\lambda}{(2\pi)^2} ~.
\label{eq:}
\end{equation}
\begin{equation}
(\omega\pm E_p)^2-E_{p-q}^2 = \omega^2 - q^2 \pm 2\omega E_p + 2pq\lambda~,
\label{eq:}
\end{equation}
\begin{equation}
2pq\lambda\rightarrow\lambda~.
\label{eq:}
\end{equation}
We are now left with two angular integrals
\begin{equation}
\int_{-2pq}^{2pq} \frac{d\lambda}{\omega^2-q^2\pm2\omega E_p+\lambda}
= 
\ln\frac{\omega^2-q^2\pm2\omega E_p+2pq}{\omega^2-q^2\pm2\omega E_p-2pq}~,
\label{eq:}
\end{equation}
\begin{equation}
\int_{-2pq}^{2pq} \frac{\lambda d\lambda}{\omega^2-q^2\pm2\omega E_p+\lambda}
= 
4pq - [\omega^2-q^2\pm2\omega E_p]\ln\frac{\omega^2-q^2\pm2\omega E_p+2pq}{\omega^2-q^2\pm2\omega E_p-2pq}~,
\label{eq:}
\end{equation}
\begin{equation}
R_\pm(\omega) = \omega^2-q^2-2\omega E_p\pm2pq~.
\label{eq:}
\end{equation}
Using the definition $\textrm{Re}f(\omega)=\frac{1}{2}[f(\omega)+f(-\omega)]$
we obtain for the longitudinal component of (\ref{eq:Decomposition}) the result
\begin{equation}
F(\omega,q) = -g^2\frac{\omega^2-q^2}{q^2}\textrm{Re}\int_0^\infty \frac{p^2dp}{\pi^2} \frac{2f_\Phi(E_p)}{E_p}
\Bigg\{1 + \frac{4E_p\omega + q^2-\omega^2-4E_p^2}{4pq}\ln\frac{R_+(\omega)}{R_-(\omega)}\Bigg\}~.
\label{eq:}
\end{equation}
To get the electric screening mass (Debye mass)
we have to compute $F(0,q\rightarrow0)=m_D^2$, i.e., 
\begin{equation}
F(0,q) = g^2\int_0^\infty \frac{p^2dp}{\pi^2} \frac{f_\Phi(E_p)}{E_p}
\Bigg\{2 + \frac{q^2-4E_p^2}{4pq}\ln\frac{(2p-q)^2}{(2p+q)^2}\Bigg\}~, 
\label{eq:}
\end{equation}
\begin{equation}
\lim_{q\rightarrow0}\frac{1}{q}\ln\frac{(2p-q)^2}{(2p+q)^2} = - \frac{2}{p}~,
\label{eq:}
\end{equation}
\begin{equation}
F(0,q\rightarrow0) = m_D^2 = g^2\int_0^\infty \frac{dp}{\pi^2}\frac{2f_\Phi(E_p)}{E_p}
\left\{p^2 + E_p^2\right\}~.
\label{eq:}
\end{equation}
This result has a structure exactly as the QED Debye mass
with the only difference that the Fermi-Dirac distribution 
function is replaced with the Polyakov loop suppressed distribution.
After including the factors $N_f=2$ and $\alpha_s=g^2/4\pi$
we get
\begin{equation}
m_D^2 = \frac{16\alpha_s}{\pi}\int_0^\infty~dp\frac{f_\Phi(E_p)}{E_p}
\left\{p^2 + E_p^2\right\}~,
\label{eq:}
\end{equation} 
which is the formula (\ref{debyemass}) from the text.
\end{appendix}



\end{document}